\documentstyle[12pt,epsfig]{article}
\advance \topmargin by -\headheight
\advance \topmargin by -\headsep
\evensidemargin \oddsidemargin
\begin{document}
\pagestyle{plain}
\begin{titlepage}
\vspace*{0.15cm}
\begin{center}
{\large\bf INSTITUTE FOR HIGH ENERGY PHYSICS}
\end{center}
\vspace*{1.cm}

\hfill {\bf IHEP-2003-18}

\vspace*{3.cm}

\begin{center}
{\Large\bf
    High statistics study of the $K^{-} \rightarrow \pi^{0} e^{-} \nu $ 
    decay} \\

\vspace*{0.15cm}
\vspace*{1.3cm}

{\bf  I.V.~Ajinenko, S.A.~Akimenko,  K.S.~Belous, 
 G.I.~Britvich, I.G.Britvich, K.V.Datsko,  A.P.~Filin, 
A.V.~Inyakin,  A.S.~Konstantinov, V.F.~Konstantinov,  
I.Y.~Korolkov, V.A.~Khmelnikov, V.M.~Leontiev, V.P.~Novikov,
V.F.~Obraztsov,  V.A.~Polyakov, V.I.~Romanovsky, V.M.~Ronjin, 
   V.I.~Shelikhov, N.E.~Smirnov, A.A.~Sokolov, 
  O.G.~Tchikilev, V.A.Uvarov, O.P.~Yushchenko. }
\vskip 0.15cm
{\large\bf $Institute~for~High~Energy~Physics,~Protvino,~Russia$}
\vskip 0.35cm
{\bf V.N.~Bolotov, S.V.~Laptev, A.R.~Pastsjak, A.Yu.~Polyarush, 
 R.Kh.~Sirodeev. }
\vskip 0.15cm
{\large\bf $Institute~for~Nuclear~Research,~Moscow,~Russia$}
\end{center}

\vspace*{3cm}

\begin{abstract}
 The  decay $K^{-} \rightarrow \pi^{0} e \nu$ has been 
 studied using in-flight decays detected with the "ISTRA+" spectrometer 
 operating 
 in the 25-GeV negative secondary beam of the U-70 PS. About 550K events were
 collected for the analysis. The $\lambda_{+}$ parameter of the vector 
 form-factor has been measured:
 $\lambda_{+}= 0.0286 \pm 0.0008 (stat) \pm 0.0006(syst)$. 
 The limits on the 
 possible tensor and scalar couplings have been obtained:
 $f_{T}/f_{+}(0)=0.021^{+0.064}_{-0.075} (stat) \pm 0.026(syst) ; $ 
 $f_{S}/f_{+}(0)=0.002^{+0.020}_{-0.022}(stat) \pm 0.003(syst) $
\end{abstract}

\end{titlepage}

\newpage
\thispagestyle{empty}

~

\setcounter{page}{0}
\newpage
\raggedbottom
\sloppy

\section{ Introduction}
 
 The decay $K \rightarrow e \nu \pi^{0}$(K$_{e3}$) is known to be a very 
 promising one to 
 search for an  admixture of scalar (S) or tensor (T) contributions to the
 Standard Model (SM) V-A amplitude. This decay 
 has been extensively studied over recent years
 in different experiments with the charged and neutral kaons. 
 Some published results indicate the anomalous S and T 
 signals \cite{Akim1,KTeV}.

 On the other hand, a recent KEK 
 experiment \cite{KEK1,KEK2}
 with a stopped $K^{+}$ beam and our preliminary studies 
 \cite{Paper1} do not observe any visible contributions of tensor and scalar
 interactions.
 Another goal of our study is a precise measurement of the V-A
 $f_{+}(t)$ form-factor of the  $K_{e3}$ decay which is interesting 
 in view of new
 calculations in the Chiral Perturbation Theory(ChPT) up to 
 the order $p^6$ \cite{Bijnens}.
   
 We present a new study of the  $K^{-}_{e3}$ decay based on the statistics    
 of about 550K  events. This anaysis is an update of our preliminary 
 results  \cite{Paper1}.
  
\section{ Experimental setup}
The experiment has been performed at the IHEP 70 GeV proton synchrotron U-70.
The experimental setup "ISTRA+" (Fig.1) 
was described in some details in our paper  \cite{Paper1}. 

\begin{figure}[h]
\epsfig{file=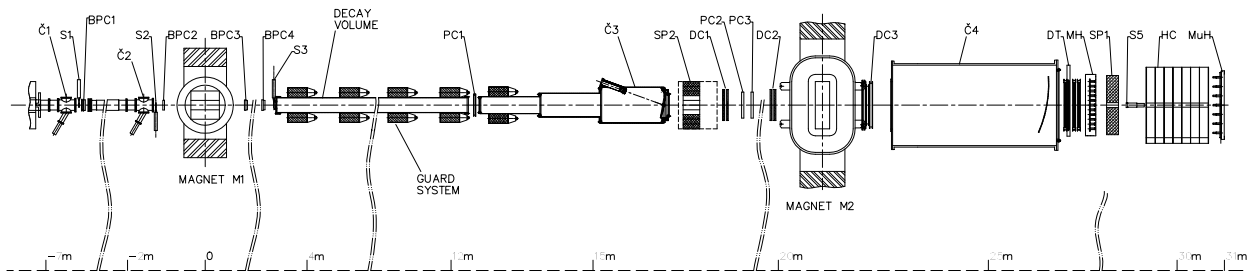, angle=90 ,width=16.5cm }
\caption{ Elevation view of the   "ISTRA+" detector.}
\end{figure}


 The setup is located in a negative unseparated secondary beam. 
The beam momentum is $\sim 25$ GeV with 
$\Delta p/p \sim 1.5 \% $. The admixture of $K^{-}$ in the beam is $\sim 3 \%$.
The beam intensity is $\sim 3 \cdot 10^{6}$ per 1.9 sec. of the U-70 spill.
The  beam particles are deflected by the beam magnet M$_{1}$ and are 
measured by $BPC_{1}\div BPC_{4}$ proportinal chambers
with 1~mm wire spacing. The kaon identification is performed by 
$\check{C_{0}} \div \check{C_{2}}$ threshold $\check{C}$-counters. 

The 9 meter long vacuumed
decay volume is surrounded by 8 lead-glass rings $LG_{1} \div LG_{8}$ 
which are used as the veto system for 
low energy photons. The photons radiated at large angles are detected 
by the lead-glass calorimeter $SP_{2}$.

The decay products are deflected by the spectrometer magnet  M2 with 
a field integral of 1~Tm. The track measurement is performed by 
 2-mm-step proportional chambers  
($PC_{1} \div PC_{3}$), 1-cm-cell drift chambers  
($DC_{1} \div DC_{3}$),
 and by 2-cm-diameter
drift  tubes    ($DT_{1} \div DT_{4}$). 
Wide aperture threshold \v{C}erenkov counters ($\check{C_{3}}$ and 
$\check{C_{4}}$) are filled  with helium and
are not used in these measurements. 

The photons are measured by the lead-glass calorimeter $SP_{1}$ which consists
of 576 counters. The counter transverse  
size is $5.2\times 5.2$ cm and the 
length is about 15~$X_0$. 

The  scintillator-iron sampling hadron calorimeter HC is subdivided
into 7 longitudinal sections 7$\times$7 cells each. The 11$\times$11 cell 
scintillating hodoscope is used for the improvement of the time
resolution of the tracking system. 
MuH  is a 7$\times$7 cell scintillating muon hodoscope.

The trigger is provided by $S_{1} \div S_{5}$ scintillation counters, 
$\check{C_{0}} \div \check{C_{2}}$ Cerenkov counters, and
the analog sum of amplitudes from last dinodes of the $SP_1$ :
\begin{displaymath}
 T=S_{1} \cdot S_{2} \cdot S_{3} \cdot 
 \bar{S_{4}} \cdot \check{C_{0}} \cdot \bar{\check{C_{1}}} \cdot 
 \bar{\check{C_{2}}} \cdot 
 \bar{S_{5}} \cdot \Sigma(SP_{1}),
\end{displaymath}
where $S_4$ is the scintillator counter with a hole to suppress the beam halo,
 $S_5$ is the counter located downstream  the setup at the beam focus,
$\Sigma(SP_{1})$ requires that the analog sum of amplitudes from 
the $SP_1$ be larger than $\sim$700 MeV - a MIP signal. The last requirement 
serves to suppress the dominating $K \rightarrow \mu \nu$ decay.

\section{Events selection}

During runs in Spring (run1) and Winter (run2) 2001, 
363M and 332M events were logged on tapes.
This statistics is supported by about 260M MC events generated with 
Geant3 \cite{geant} Monte Carlo program. The MC generation
includes a realistic description of the setup with decay volume 
entrance windows,
tracking chambers windows, chambers gas mixtures, sense wires and cathode 
structures,
\v{C}erenkov counters mirrors and gas, the shower generation in EM calorimeters, 
etc.

\begin{minipage}[t]{8.cm}
\epsfig{file=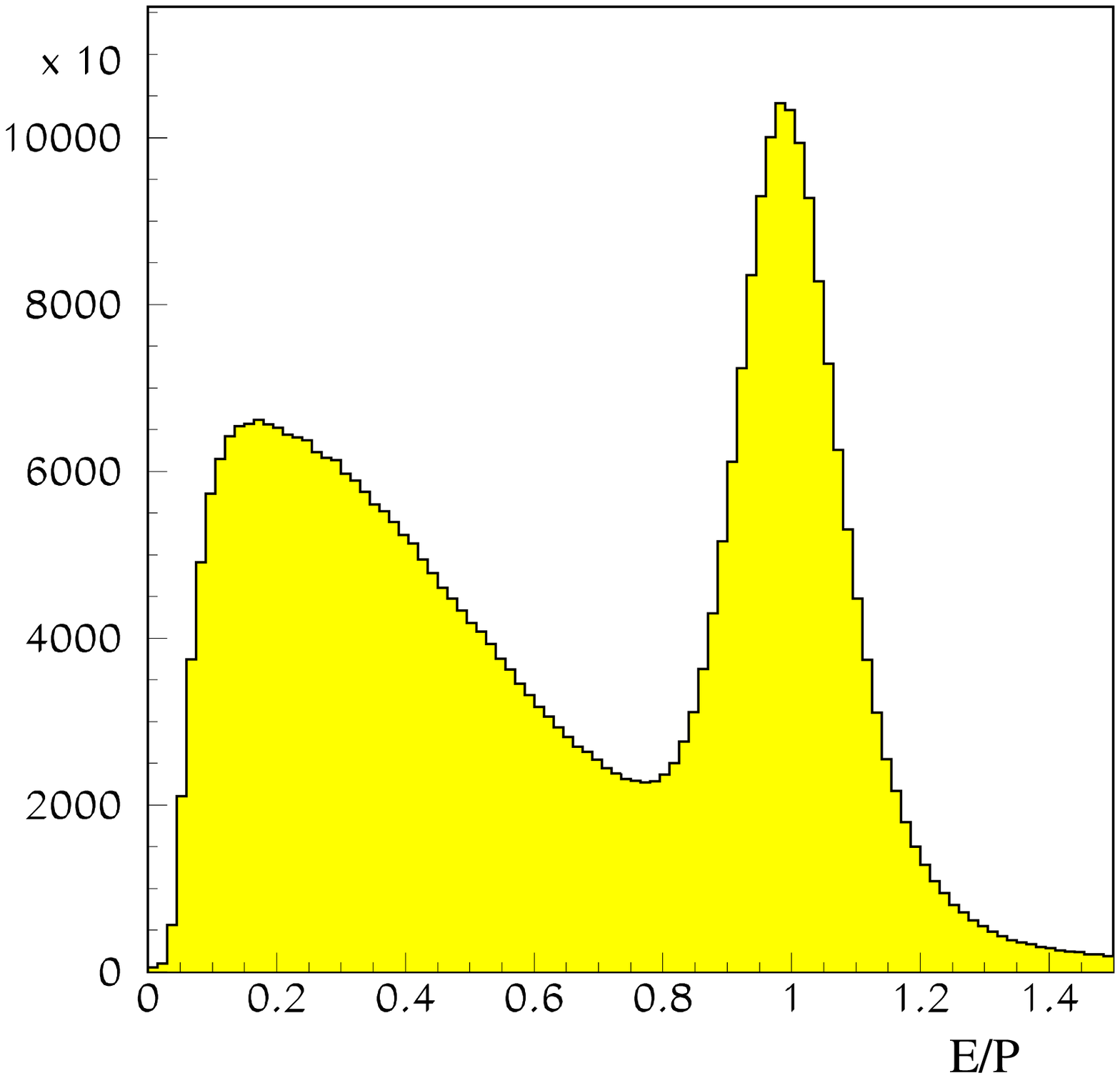,width=8cm}

\begin{center}
Figure 2: The ratio of the energy of the associated ECAL
 cluster to the momentum of the charged track.
\end{center}
\end{minipage} \ \hfill \ 
\begin{minipage}[t]{8.cm}
\epsfig{file=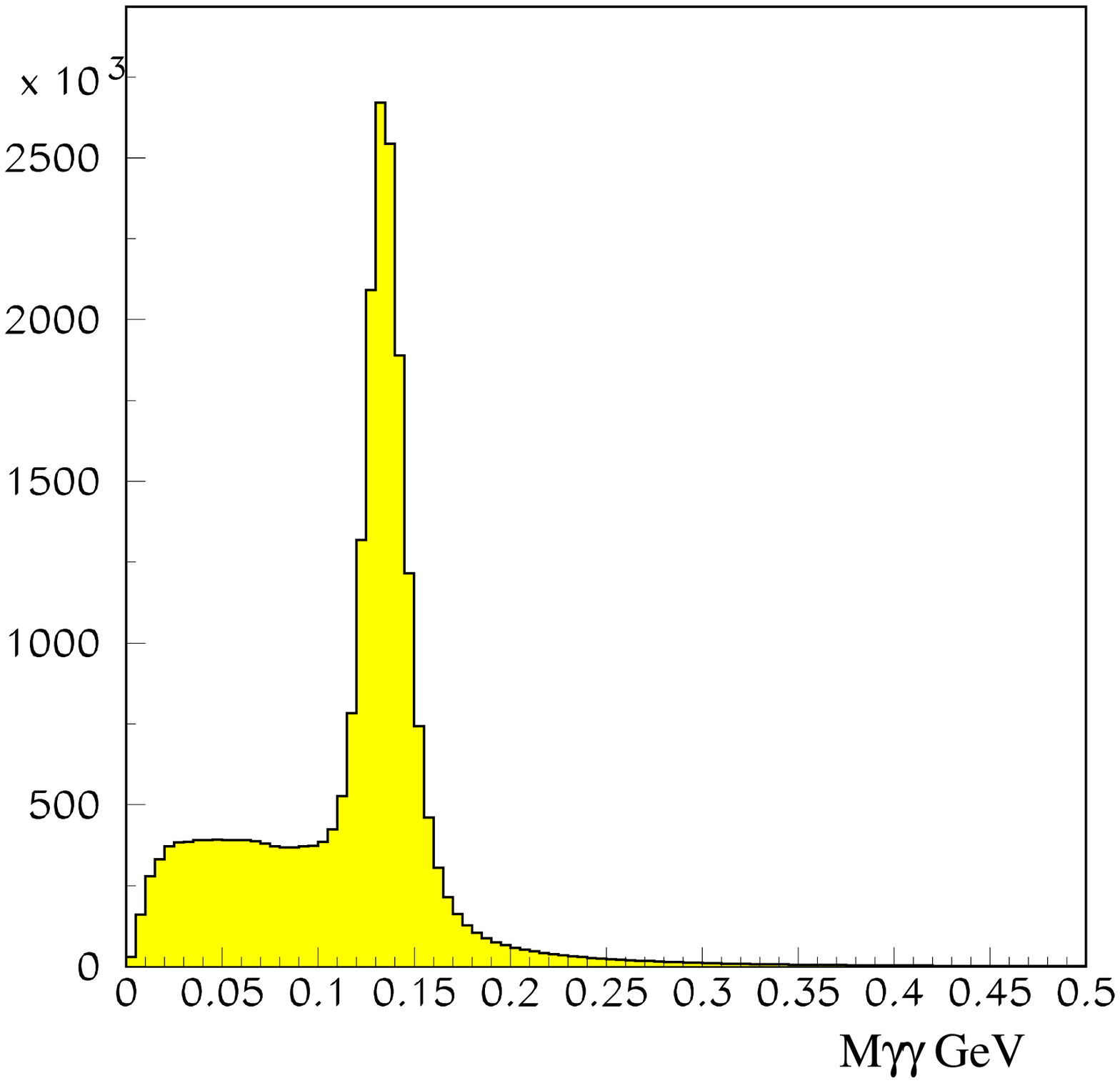,width=8cm}
\begin{center}
Figure 3: The $ \gamma \gamma$ mass spectrum for the events with
 the identified electron and two extra showers.
\end{center}
\end{minipage}

~

~

The data processing starts with the beam particle reconstruction in 
$BPC_{1} \div BPC_{4}$. Then secondary tracks are looked for in the
decay tracking system
and events with one good negatively charged track are selected.
The decay vertex is reconstructed by means of the 
unconstrained vertex fit of the beam and decay tracks.

A clustering procedure is used to find showers in the 
$SP_{1}$ calorimeter, and the two-dimensional pattern of the shower is fitted 
with the MC-generated patterns to 
reconstruct its energy and position.

  The matching of the charged track and a shower 
in $SP_{1}$ is done 
on the basis of the distance r between the track extrapolation to the 
calorimeter and the 
shower coordinates ($r\leq 3$cm). The electron identification is done using 
the ratio of the shower energy (E)
to the momentum of the associated track (P). 
The E/p distribution is shown in 
Fig.2. 
The particles
with $0.8 < \mbox{E/p} < 1.3$ are accepted as electrons.

 The events with one charged track identified as electron and two additional 
showers in the ECAL are selected for further processing. 
The $\gamma \gamma$-mass spectrum is shown in Fig.3.
The $\pi^{0}$ peak is situated at 
$M_{\pi0}=134.8$ MeV with a resolution of 8.6 MeV.

 The selected events are required to pass 2C
 $K \rightarrow e \nu \pi^{0}$ fit.
 To  minimize effects of a beam associated background and systematics  
the cut on the difference between
the fitted $P_{K}$ value and the mean beam energy (25.2 GeV for the first run
and 26.3 GeV for the second) is applied: $|P_{K}-\bar{P}_{beam}|<1$ GeV
(see Fig.4).
  The missing energy 
 $E_{\nu}=E_{K}-E_{e}-E_{\pi^{0}}$ at this stage of the selection
is shown in Fig.5 for run2 data. 

\begin{minipage}[t]{8.cm}
\epsfig{file=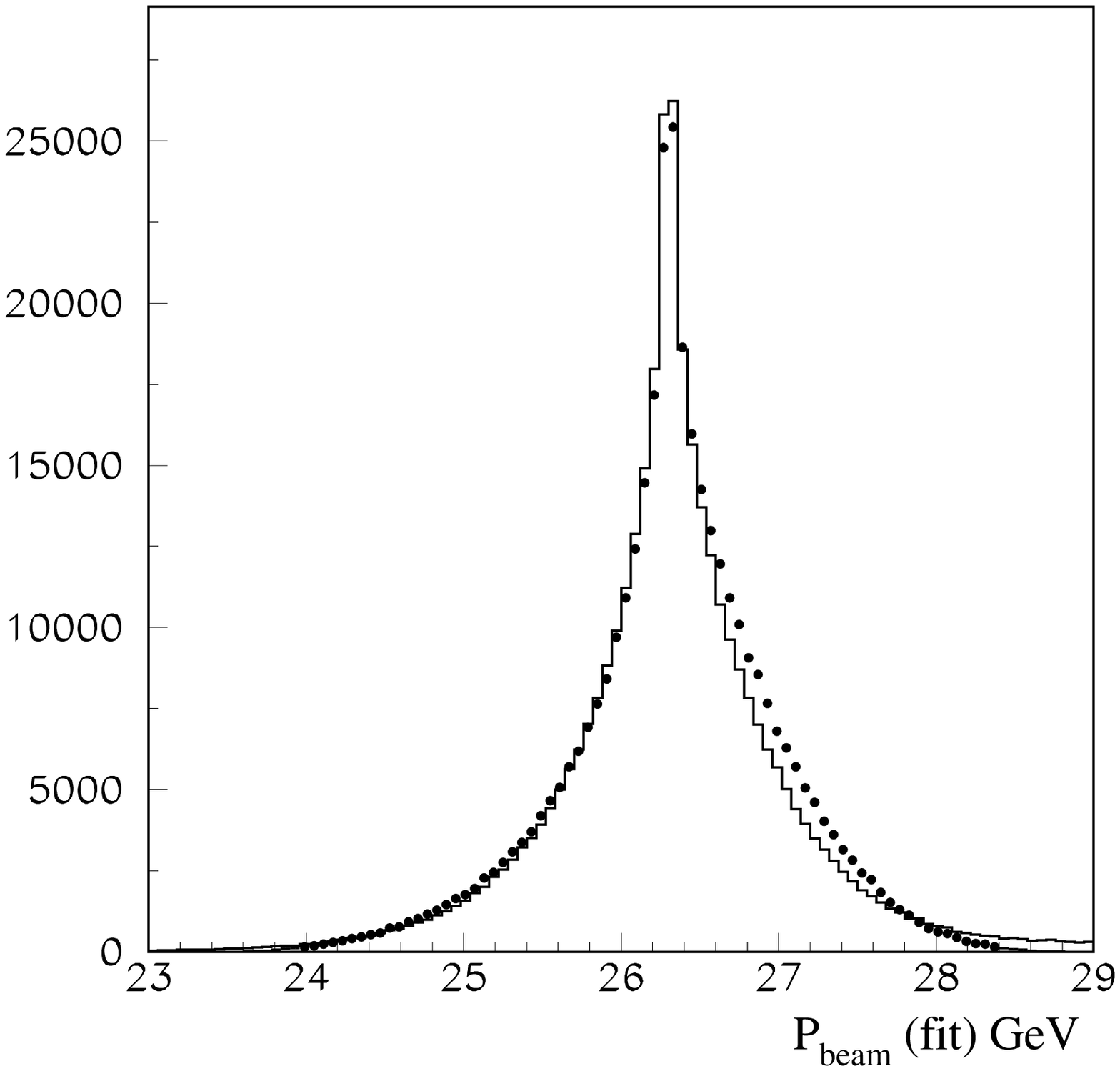,width=8cm}

\begin{center}
Figure 4: The beam kaon momentum after 2C $K_{e3}$ fit.
 The points with errors are the run2 data and the histogram is MC. 
\end{center}
\end{minipage} \ \hfill \ 
\begin{minipage}[t]{8.cm}
\epsfig{file=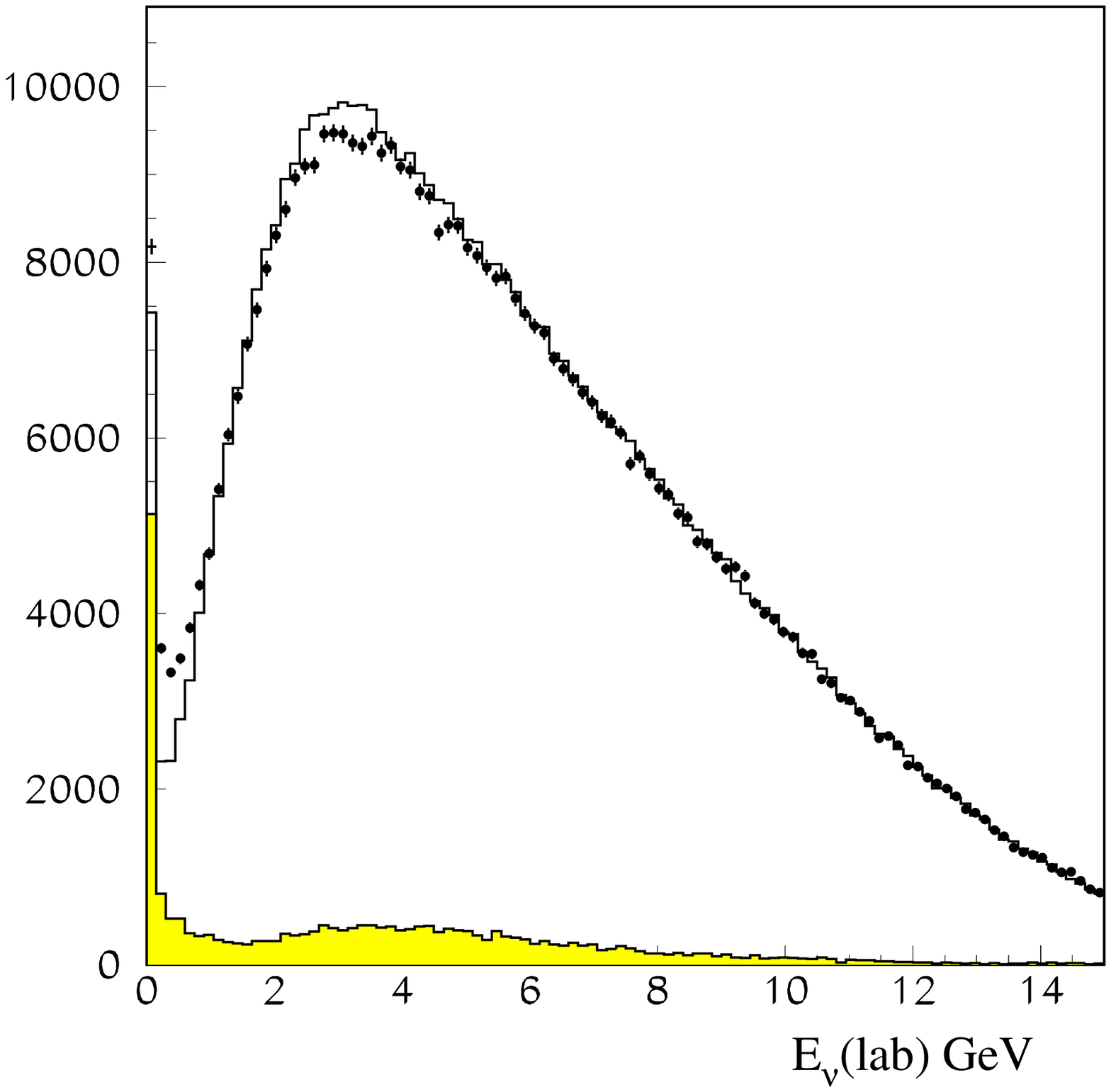,width=8cm}
\begin{center}
Figure 5: The $E_\nu$ energy compared with MC.
The points are data, the dark histogram is the
 background contamination.
\end{center}
\end{minipage}

~

~

The peak at low $E_{\nu}$ corresponds to
the remaining $K^{-} \rightarrow \pi^{-} \pi^{0}$ background. 
The cut $E_{\nu}>0.3$ GeV is applied to suppress this background. 
The requirement that the 5C $K \rightarrow \pi^{-} \pi^{0}$ fit 
should fail is applied for further background suppression.

 The surviving background is estimated to be less
than $1.5\%$.

\section{ Analysis}

Performing the procedure described in the previous section,
112K and 440K events are selected in run1 and run2 data. 
The difference in the event output
is explained by the higher $SP_1$ analog sum threshold in the second run.
The distribution of
the events over the Dalitz plot for the run2 is shown in Fig.6. The  variables 
$y=2E_{e}/M_{K}$ and $z=2E_{\pi}/M_{K}$, where $E_{e}$ and $E_{\pi}$ are the
energies of the electron and $\pi^{0}$ in the kaon rest frame, are used. 
The analysis of the MC-generated background
shows that the background events are
concentrated at the peripheral part of the plot.

The most general Lorentz-invariant form of the matrix element for the 
$K^{-} \rightarrow l^{-} \nu \pi^{0}$ decay is \cite{Steiner}:
\begin{equation}
M= \frac{G_{F}V_{us}}{2} \bar u(p_{\nu}) (1+ \gamma^{5})
[2m_{K}f_{S} -
[(P_{K}+P_{\pi})_{\alpha}f_{+}+
(P_{K}-P_{\pi})_{\alpha}f_{-}]\gamma^{\alpha} + i \frac{2f_{T}}{m_{K}}
\sigma_{\alpha \beta}P^{\alpha}_{K}P^{\beta}_{\pi}]v(p_{l})
\end{equation}
It consists of scalar, vector, and tensor terms. The $f_{\pm}$ form-factors
are the functions of $t= (P_{K}-P_{\pi})^{2}$. In the Standard Model (SM),
the W-boson exchange leads to the pure vector term. 
The scalar and/or tensor terms which are ``induced'' by
 EW radiative corrections are negligibly
small, i.e  nonzero scalar or tensor form-factors would indicate 
the physics beyond the SM. 

The term in the vector part, proportional to $f_{-}$, 
is reduced (using the Dirac
equation) to the scalar form-factor. In the same way, the tensor term is 
reduced to
a mixture of the scalar and  vector form-factors. The redefined vector (V) and 
scalar (S) terms, and the corresponding Dalitz plot
density in the kaon rest frame ($\rho(E_{\pi},E_{l})$) are \cite{Chizov}:
\begin{eqnarray}
\rho (E_{\pi},E_{l}) & \sim & A \cdot |V|^{2}+B \cdot Re(V^{*}S)+C \cdot |S|^{2} \\
 V & = & f_{+}+(m_{l}/m_{K})f_{T} \nonumber \\ 
S & = & f_{S} +(m_{l}/2m_{K})f_{-}+
\left( 1+\frac{m_{l}^{2}}{2m_{K}^{2}}-\frac{2E_{l}}{m_{K}}
-\frac{E_{\pi}}{m_{K}}\right) f_{T} \nonumber \\ 
A & = & m_{K}(2E_{l}E_{\nu}-m_{K} \Delta E_{\pi})-  
m_{l}^{2}(E_{\nu}-\frac{1}{4} \Delta E_{\pi}) \nonumber \\
B & = & m_{\l}m_{K}(2E_{\nu}-\Delta E_{\pi}) ;~ E_{\nu}=m_{K}- E_{l}-E_{\pi}
 \nonumber \\
C & = & m_{K}^{2} \Delta E_{\pi};~ \Delta E_{\pi}  =  E_{\pi}^{max}-E_{\pi} ;~
E_{\pi}^{max}= \frac{m_{K}^{2}-m_{l}^{2}+m_{\pi}^{2}}{2m_{K}} \nonumber 
\end{eqnarray}
 The terms proportional to 
$m_{l}$ and $m_{l}^{2}$ can be neglected in the case of the $\mbox{K}_{e3}$ 
decay.

Then, assuming the linear dependence of the $f_{+}$
on t: $f_{+}(t)=f_{+}(0)(1+\lambda_{+}t/m_{\pi}^{2})$ and real constants 
 $f_{S}$, $f_{T}$,  we get:  
\begin{eqnarray}
\rho (E_{\pi},E_{l}) & \sim &
 m_{K}(2E_{l}E_{\nu}-m_{K} \Delta E_{\pi})\cdot
  (1+\lambda_{+}t/m_{\pi}^{2})^{2} \nonumber \\
 & + & m_{K}^{2} \Delta E_{\pi}\cdot \left( \frac{f_{S}}{f_{+}(0)} +
\left( 1-\frac{2E_{l}}{m_{K}}
-\frac{E_{\pi}}{m_{K}}\right) \frac{f_{T}}{f_{+}(0)}\right)^{2}     
\end{eqnarray}

The procedure of the extraction of the parameters
$ \lambda_{+}$, $f_{S}$, $f_{T}$ starts with the subdivision of the 
  Dalitz plot region 
$ y= 0.12 \div 0.92;\; z=0.55 \div 1.075$  into $40\times 40$ cells.

The number of events in a cell (i,j) of the Dalitz plot, 
in the case of the simultaneous extraction of, for example, 
$\lambda_{+}$ and $\frac{f_{S}}{f_{+}(0)}$,
 is fitted by the function:
\begin{eqnarray}
W^{MC} (i,j)\sim W_{1}(i,j)+W_{2}(i,j) \cdot \lambda_{+}+
W_{3}(i,j) \cdot \lambda_{+}^{2}+ W_{4}(i,j) 
\cdot \left( \frac{f_{S}}{f_{+}(0)}\right)^{2}      
\end{eqnarray} 
Here $W_{l}$ are MC-generated  functions which are built up as follows:
the MC events are generated with the constant density over the Dalitz plot 
and 
reconstructed with the same program as for the real events.  Each event
carries the weight $w$ determined by the corresponding term in the 
expression 3,  
calculated using the MC-generated ``true'' values for $y$ and $z$.  
The radiative corrections  according to \cite{grinb} are taken into account.
Then, $W_{l}$ are calculated by summing up the weights of the reconstructed 
events in
the corresponding Dalitz plot cell. This procedure allows one to avoid the
systematic errors due to the ``migration'' of the events over the Dalitz plot
due to the finite experimental resolution.

\begin{minipage}[t]{8.cm}
\epsfig{file=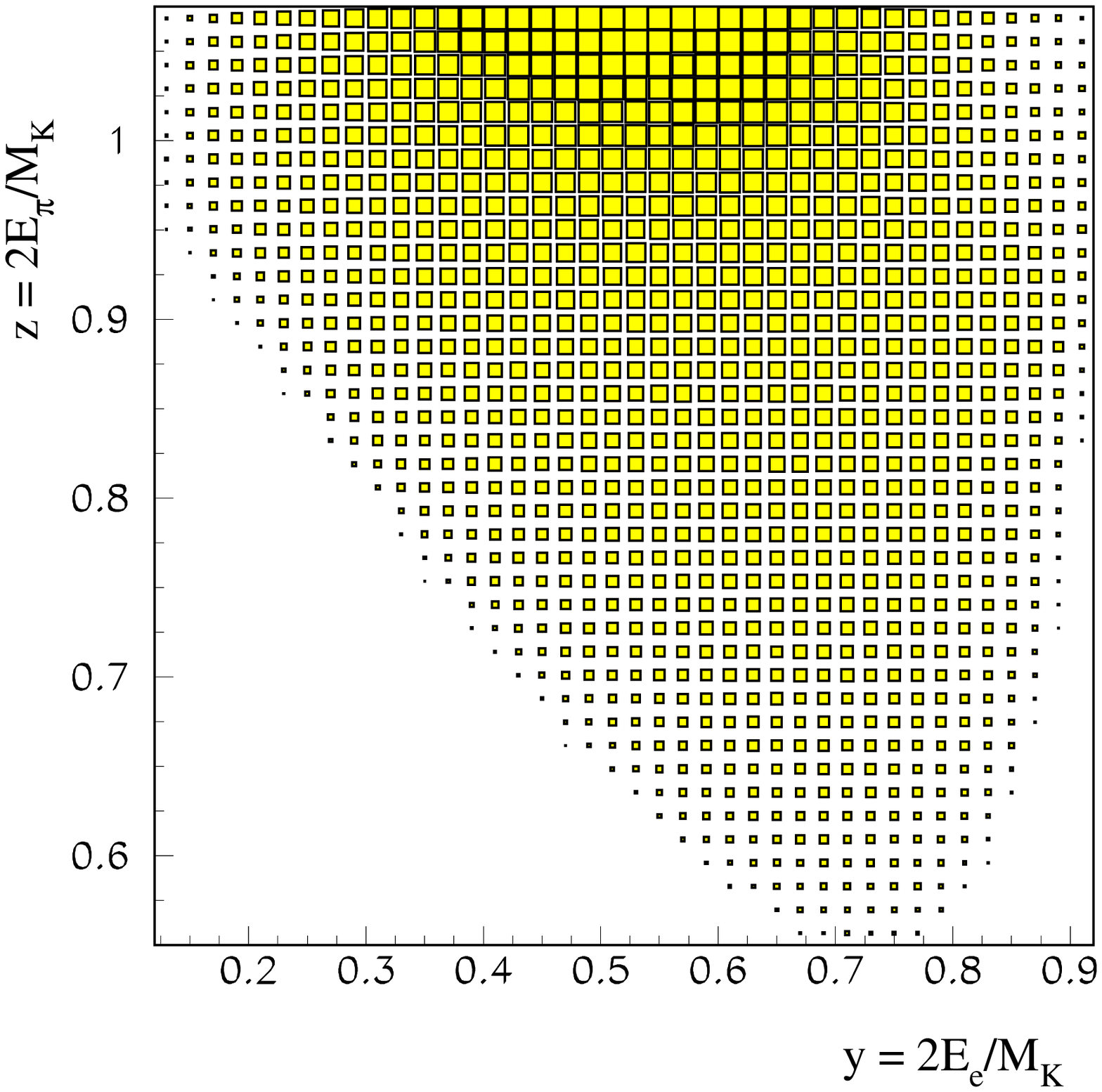,width=8cm}

\begin{center}
Figure 6: Dalitz plot for the selected \\
$K \rightarrow e \nu \pi^{0}$ events. Run2 data.
\end{center}
\end{minipage} \ \hfill \ 
\begin{minipage}[t]{8.cm}
\epsfig{file=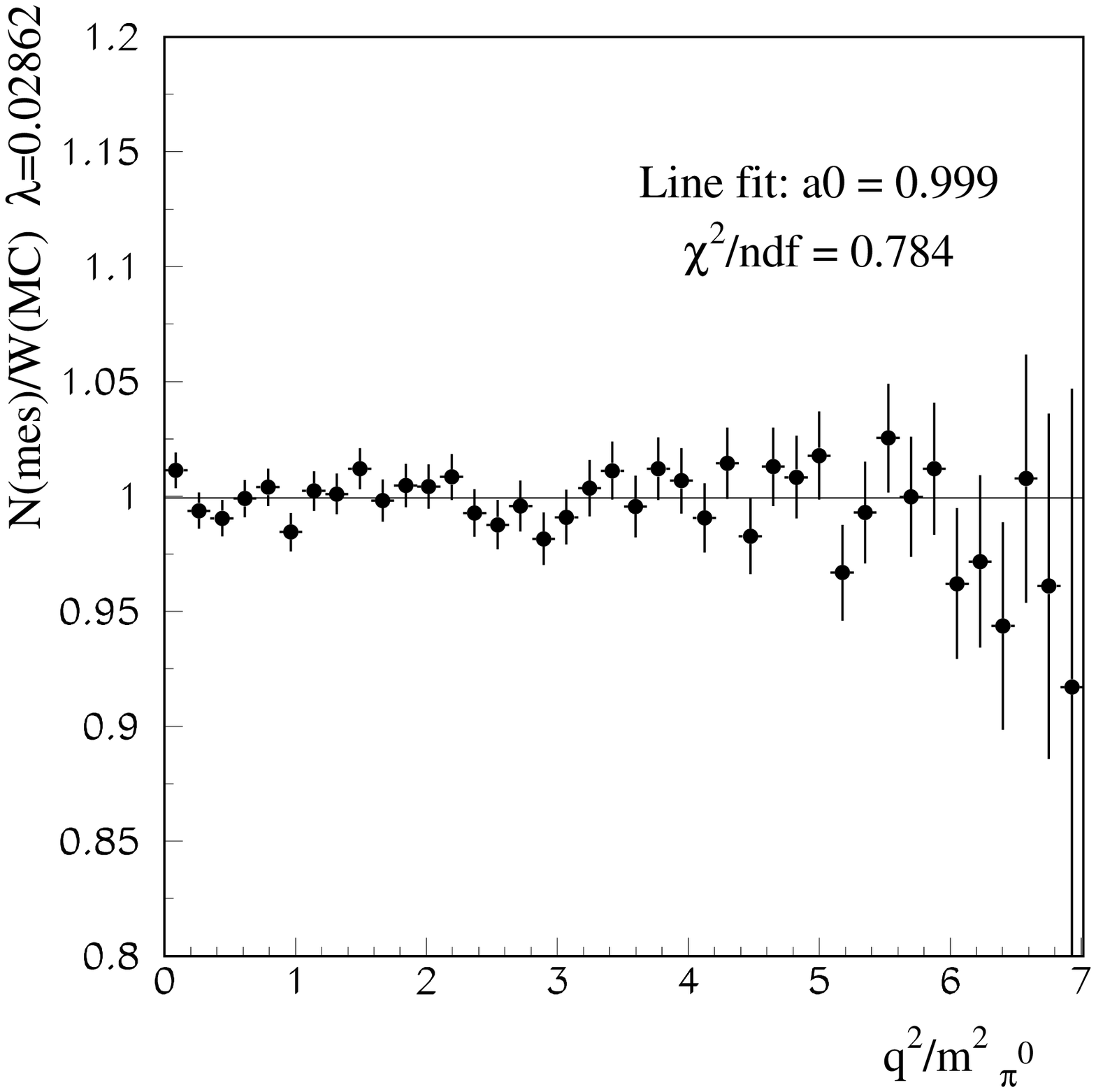,width=8cm}

\begin{center}
Figure 7: The ratio $(N^{\mbox{rd}}-N^{\mbox{bg}})/W_{MC}$ (see text) for the
 run1+run2 data.
\end{center}
\end{minipage}

~

~

The parameters of the amplitude of the decay are extracted by minimizing
the extended log-likelihood function:
\begin{equation}
   {\cal L} =  \sum_{i,j}\left\{ \left( A\cdot W_{i,j}^{MC} + 
   N^{\mbox{bg}}_{ij}- 
                              N_{i,j}^{\mbox{rd}} \right) 
   +   N_{i,j}^{\mbox{rd}}\cdot \ln 
     \frac{ N_{i,j}^{\mbox{rd}}}{A\cdot W_{i,j}^{MC}+
      N^{\mbox{bg}}_{ij}}\right\},
\end{equation}
where the sum runs over all populated cells of the Dalitz plot. Here ``A''
 is the 
data/MC  normalization parameter, $N_{ij}^{\mbox{rd}}$ -- number of real data
events in the $(i,j)$-cell, and
$N^{\mbox{bg}}_{ij}$ - the MC
estimated background. The background 
normalization is determined by the ratio of the real and generated
$K^{-} \rightarrow \pi^{-} \pi^{0}$ events.
The minimization is performed by means of the ``MINUIT'' 
program \cite{Minuit}.

\section{Results}

The results of the fit are summarized in Table 1.

The combination of two runs is done by the simultaneous fit.
The first line corresponds to the pure V-A SM fit. In the second line
the tensor  and, in the third, the scalar terms are  added into the fit.

 The final errors are calculated by the ``MINOS'' procedure of 
the ``MINUIT''
program \cite{Minuit}.
 The errors are estimated at the $2\Delta{\cal L}$ level
of 2.3 that corresponds to the coverage probability of 68.27\%\ for joint
estimation of 2 parameters \cite{PDG}.

\renewcommand{\arraystretch}{1.4}

\begin{center}
\begin{tabular}{|c|ccc|}
\hline
    & run1  & run2 & run1+run2 \\ \hline\hline
    
  $\lambda_+$ &  $0.0291 \pm 0.0018$ & 
                 $0.0285 \pm 0.0009$  &
                 $0.0286 \pm 0.0008$ \\ \hline		 
 $f_T/f_+(0)$ &  $0.018^{+0.087}_{-0.099}$ &
                 $0.024^{+0.081}_{-0.107}$ &
                 $0.021^{+0.064}_{-0.075}$   \\ \hline			  
 $f_S/f_+(0)$ &  $0.003^{+0.024}_{-0.029}$ &
                 $0.002^{+0.019}_{-0.032}$ &
                 $0.002^{+0.020}_{-0.022}$   \\ \hline\hline

  $\chi^2/\mbox{ndf}$ &  1.13  & 1.14 & 1.14 \\  \hline
  $N_{\mbox{bins}}$ &  1120    & 1119 & 2239 \\ \hline\hline
\end{tabular}
\end{center}
\begin{center}
Table 1. Results of the fit
\end{center}

To illustrate the quality of the fit, Fig.7 shows the ratio of the real  data
  minus MC-predicted background  ($N^{\mbox{rd}}-N^{\mbox{bg}})$ 
  over $W_{MC}$ versus $t/m^{2}_{\pi^{0}}$,
calculated using expression 4, with $\lambda_{+}=0.0286$ and 
$f_{T}=f_{S}=0$.

Different sources of systematics are investigated:
\begin{itemize}
\item The Dalitz plot bining: \\ 
$\Delta \lambda_{+}= \pm 0.00006$; $\Delta f_{T}= \pm 0.003$;
 $\Delta f_{S}= \pm 0.0006$. This kind of systematics is very small
 in our case because of the fitting method (see expression 4).
\item Selection cuts  variation: 
\begin{enumerate}
\item The 5C  $K \rightarrow
\pi^{-} \pi^{0}$ fit probability cut :\\
$\Delta \lambda_{+}= \pm 0.00041$; $\Delta f_{T}= \pm 0.0144$;
 $\Delta f_{S}= \pm 0.0018$.
\item The 2C   
 $K \rightarrow e \nu \pi^{0}$ fit probability cut:\\
$\Delta \lambda_{+}= \pm 0.00025$; $\Delta f_{T}= \pm 0.008$;
 $\Delta f_{S}= \pm 0.0012$.
\item    The  missing energy ($E_{\nu}$) cut:\\
$\Delta \lambda_{+}= \pm 0.00015$; $\Delta f_{T}= \pm 0.0081$;
 $\Delta f_{S}= \pm 0.0008$.
\item The $|P_{K}-\bar{P}_{beam}|$ cut:\\
$\Delta \lambda_{+}= \pm 0.00026$; $\Delta f_{T}= \pm 0.0104$;
 $\Delta f_{S}= \pm 0.0018$.
\item The E/p electron selection cut:\\
$\Delta \lambda_{+}= \pm 0.00014$; $\Delta f_{T}= \pm 0.010$;
 $\Delta f_{S}= \pm 0.0008$.
\end{enumerate}
\item The signal MC variation:
$\Delta \lambda_{+}= \pm 0.00005$; $\Delta f_{T}= \pm 0.001$;
 $\Delta f_{S}= \pm 0.0001$.
\item The background MC variation:
$\Delta \lambda_{+}= \pm 0.00015$; $\Delta f_{T}= \pm 0.011$;
 $\Delta f_{S}= \pm 0.0003$.
\end{itemize}
From that, the total  systematics is:
$\Delta \lambda_{+}= \pm 0.0006$; $\Delta f_{T}= \pm 0.026$;
 $\Delta f_{S}= \pm 0.003$.
 
 The comparison of our results with the
most recent $K^{\pm}$ data \cite{Akim1,KEK2} shows very good agreement in the 
$\lambda_{+}$ parameter. We do not observe any visible contributions of scalar
and tensor terms in the amplitude, in agreement with the conclusions of 
\cite{KEK2,Paper1}.

Some difference ($2 \sigma$) with the ChPT $\mbox{O}(p^4)$ 
calculations for $\lambda_{+}$:  $\lambda_{+}=0.031$ \cite{Leutwyler}, 
is observed. 

In addition, a possible contribution of the quadratic term
$\lambda^{''}_{+}t^{2}/m_{\pi}^{4}$ into the $f_{+}$ form-factor is searched 
for. 
The 
value of $\lambda^{''}_{+} = -0.00042^{+0.0011}_{-0.0015}$ is obtained 
with the fixed  $\lambda_{+} = 0.0286$. If $\lambda_{+}$ is 
allowed to vary, its value is shifted to 
$\lambda_{+} = 0.02867 \pm 0.0020$ and the quadratic coefficient becomes
$\lambda^{''}_{+} = -0.002^{+0.0031}_{-0.0066}$. We conclude that
a possible quadratic contribution into the vector form-factor is compatible with
zero in both cases.

\section{Summary and conclusions}
The $K^{-}_{e3}$ decay has been studied using in-flight decays of 25 GeV 
$K^{-}$, detected by the ``ISTRA+'' magnetic spectrometer. Due to the high
statistics, adequate resolution of the detector, and good sensitivity over
all the Dalitz plot space, the  errors in the fitted parameters 
are significantly reduced
as compared with the previous measurements. 

 The $\lambda_{+}$ parameter of the vector form-factor 
 is measured to be: 
\begin{center} 
 $\lambda_{+}=0.0286 \pm 0.0008\; (stat) \pm 0.0006\; (syst)$.
\end{center}

The limit on the quadratic nonlinearity for $f_{+}(t)$  is obtained:
\begin{center} 
 $\lambda^{''}_{+}=-0.00042^{+0.0011}_{-0.0015}\; (stat) $.
\end{center}

The limits on the 
 possible tensor and scalar couplings are derived: 
\begin{center} 
 $f_{T}/f_{+}(0)=0.021 ^{+0.064}_{-0.075}\; (stat) \pm 0.026\; (syst) ; $ \\[3mm]
 $f_{S}/f_{+}(0)=0.002 ^{+0.020}_{-0.022}\; (stat) \pm 0.003\; (syst)$ 
\end{center}

\vspace*{0.5cm}

The work
is  supported by the RFBR grant N03-02-16330. \\


\begin{thebibliography}{99}
\bibitem{Akim1} S.A.~Akimenko  et al., Phys. Lett. {\bf B259}(1991), 225.
\bibitem{KTeV} R.J.~Tesarek, hep-ex/9903069, 1999.
\bibitem{KEK1} S.~Shimizu et al., Phys. Lett. {\bf B495}(2000), 33.
\bibitem{KEK2} A.S.~Levchenko et al., Yad.Fiz  {\bf v65}(2002), 2294,
hep-ex/0111048(2001).
\bibitem{Paper1} I.V.~Ajinenko et al., Yad.Fiz {\bf v65}(2002), 2125.
\bibitem{Bijnens} J. Bijnens hep-ph/0303103(2003).
\bibitem{geant} R~ Brun et al., CERN-DD/EE/84-1.
\bibitem{Steiner} H.~Steiner et al., Phys.Lett. {\bf B36}(1971), 521.
\bibitem{Chizov} M.V.~Chizhov hep-ph/9511287(1995).
\bibitem{grinb} E.S.~Grinberg, Phys. Rev. 162 (1967), 1570.
\bibitem{Minuit} F.~James, M.Roos, CERN D506,1989.
\bibitem{Leutwyler} J.~Gasser, H.~Leutwyler Nucl. Phys. {\bf B250}(1985), 517.
\bibitem{PDG} Review of Particle Physics, Phys.Rev {\bf D66}(2002), 1.
\end{thebibliography}
\end{document}